\documentclass[11pt,a4paper]{article}

\usepackage[utf8]{inputenc}
\usepackage{amsmath}
\usepackage[mathcal]{euscript}
\usepackage{latexsym}
\usepackage{textcomp}
\usepackage{slashed}
\usepackage{tikz}
\usepackage{xspace}
\usepackage{setspace}
\usepackage[a4paper]{geometry}
\usepackage[numbers,sort&compress]{natbib}
\newcommand{\eprint}{}
\newcommand{\beq}{\begin{equation}}
\newcommand{\eeq}{\end{equation}}
\newcommand{\bqa}{\begin{eqnarray}}
\newcommand{\eqa}{\end{eqnarray}}
\newcommand{\bite}{\begin{itemize}}
\newcommand{\eite}{\end{itemize}}
\newcommand{\bd}{\begin{displaymath}}
\newcommand{\ed}{\end{displaymath}}
\def\db#1{\bar D_{#1}}

\def\Formcalc{{{\sc FormCalc}}}

\def\Gosam{{{\sc GoSam}}}

\def\gosam{{{\sc GoSam}}}

\def\blackhat{{{\sc BlackHat}}}
\def\openloops{{{\sc OpenLoops}}}
\def\madloop{{{\sc MadLoop}}}

\def\helacnlo{{{\sc Helac-NLO}}}

\def\recola{{{\sc recola}}}

\def\njet{{{\sc NJet}}}
\def\rocket{{{\sc Rocket }}}
\def\samurai{{{\sc samurai}}}
\def\SAMURAI{{{\sc samurai}}}

\def\cuttools{{{\sc CutTools}}}
\def\C++{{{\sc c++}}}

\def\Golem{{{\sc Golem95C}}}

\def\Ninja{{{\sc Ninja}}}

\begin{document}

\begin{center}

{\Large\bf Automated higher-order calculations: Status and prospects} \\
\vskip 10mm

{\bf Giovanni Ossola}\\
\small{ \tt gossola@citytech.cuny.edu}
\\[1em]
{\small {\it  New York City College of Technology, 
  City University of New York, \\ 300 Jay Street, Brooklyn NY 11201, USA} \\

  \vspace{0.2cm}

{\it The Graduate School and University Center, 
  City University of New York,  \\ 
  365 Fifth Avenue, New York, NY 10016, USA}
}
\end{center}

\vspace{0.5cm}

\begin{abstract}
\noindent  In this presentation we review the current status in the automated evaluation of scattering amplitudes, with particular attention to the developments related with NLO calculations, which led to the construction of powerful multi-purpose computational tools. 
After a general overview, we will devote a short section to describe the \gosam\ framework for NLO calculations and its application to the production of Higgs boson plus jets. We will then briefly comment  on the challenges presented by NNLO calculations, whose structure is considerably more complicated. Finally, we will describe some of the features of the integrand-reduction techniques beyond NLO, an alternative promising approach to multi-loop calculations which is currently under development.
\end{abstract}

\vspace{0.2cm}

\begin{center}
{\it to appear in the proceedings of the} \\ XXIII International Workshop on Deep-Inelastic Scattering and Related Subjects (DIS2015)\\
 April 27 - May 1, 2015\\ Dallas, Texas, USA
\end{center}

\vspace{0.3cm}

\section{Introduction}

The evaluation of scattering amplitudes allows us to test the theoretical models and compare their phenomenological predictions with the results of the experiments at particle colliders. In the light of the ongoing activities at the Large Hadron Collider (LHC), it is mandatory to have precise and reliable tools, that allow for an accurate and efficient evaluation of cross sections and differential distributions for a variety of processes. 

In the past decade, a better understanding of the structure of scattering amplitudes was achieved thanks to the complementary work of several groups, which transformed beautiful mathematical properties of scattering amplitudes, such as recursion relations, unitarity, and integrand decomposition, into practical computational tools for the evaluation of physical observables.

In order to properly describe the data collected by the experimental collaborations, theory predictions are not reliable without accounting for higher orders, since leading-order (LO) results, usually obtained with a tree-level calculation, are affected by large theoretical errors. For most analyses, results should be provided at least at Next-to-Leading-Order precision (NLO), which are considerably more involved:  they require the computation of one-loop virtual corrections (virtual part), contributions from real emission (real part), obtained by adding one additional particle in the final states, as well as a clever way of dealing with infrared divergences that occur in both virtual and real part and only cancel out when all parts are combined together. While the LO matrix elements and the NLO real parts have been available for a long time, until recently the evaluation of the virtual part of one-loop contributions represented the bottleneck towards the automation of NLO calculations.
This is not the case anymore. 

The scope of this talk is to summarize the recent progress in the evaluation and automation of scattering amplitudes, 
which led to the development of powerful automated computational tools for Next-to-Leading Order (NLO) 
calculations. After an overview of the many different tools which are currently available, we will devote a short paragraph to describe the \gosam\ framework for NLO calculations and its application to the production of Higgs boson plus jets. 

While the NLO tools are reaching their full maturity, and they are seamlessly being incorporated in the Monte Carlo programs or used to produce N-tuples of events to be used within the experimental analyses, the attention of the theoretical community is quickly shifting towards the new challenges presented by Next-to-Next-to-Leading-Order (NNLO) calculations, whose structure is considerably more complicated.
We will touch on this topic during the second part of the presentation. As a development potentially relevant for future calculations beyond NLO, we will describe the extensions of the integrand reduction to higher orders in perturbation theory. 

All the topics contained in this brief talk are based on a rich and extensive literature. We refer the reader to Refs.~\cite{review} (and references therein) for a more comprehensive picture of the different aspects of the field.   

\section{Scattering Amplitudes at Next-to-Leading Order}
\label{nlo}

The standard method for the evaluation of NLO virtual corrections relies on the calculation of all
the Feynman integrals associated with each process, namely to compute, for each diagram contributing to the amplitude and for each phase space point, integrals of the kind
{\small \beq \label{integr}
{\cal M} = \int d^n {\bar q} \,\, {\cal A} (\bar q) = \int d^n {\bar q} \frac{{\cal N}(\bar q)}{\db{0} \db{1} \ldots \db{m-1}}\, ,
\eeq}
where $\db{i}$ are the $d-$dimensional denominators generated by the propagators of the particles inside the loop.
Since any one-loop integral ${\cal M}$  can be decomposed in terms of a finite and known set of scalar master integrals (MIs)~\cite{Passarino:1978jh}, plus an additional term ${\cal R}$ known in the literature as \emph{rational part},
the calculation of one-loop virtual amplitudes can be summarized in terms of three separate tasks:
i) the \emph{generation} of the unintegrated amplitudes ${\cal A}$, namely their numerator functions  ${\cal N}$ and the list of denominators $\db{i}$;
ii) the \emph{reduction} of the amplitude to determine all coefficients multiplying each of the MIs and the rational term ${\cal R}$; 
iii) the \emph{evaluation of the MIs} which, multiplied by the coefficients obtained in the reduction, provide the final result. 
Since all scalar master integrals are known and available in public codes~\cite{scalars}
and amplitudes can be easily generated with algebraic or numerical techniques,  the reduction step is what usually distinguishes the different tools available on the market.

During the past decade, the approach to one-loop calculations was revolutionized by merging the idea of four-dimensional unitarity-cuts~\cite{Bern:1994zx,Britto:2004nc}, with the understanding of the universal algebraic form of any one-loop integrals in four dimensions, contained in the OPP method~\cite{Ossola:2006us}. Unitarity-based methods and integrand-level reduction techniques provided the theoretical background for  development of efficient computational algorithms for NLO calculations in perturbation theory, which have been implemented in various automated codes.  Tools based on generalized unitarity methods, such as \rocket~\cite{Giele:2008bc}, \blackhat~\cite{Berger:2008sj}, and  \njet~\cite{Badger:2012pg} have been very effective in tackling processes with high multiplicities, such as $W,Z+$jets or multi gluon amplitudes. The traditional $4-$dimensional OPP integrand reduction, implemented in the code {\cuttools}~\cite{Ossola:2007ax} and utilized by \madloop~\cite{Hirschi:2011pa} and  \helacnlo~\cite{Bevilacqua:2011xh}, as well as the $d-$dimensional integrand reduction provided by {\samurai}~\cite{Mastrolia:2010nb} and the integrand reduction via Laurent expansion~\cite{Mastrolia:2012bu} implemented in {\Ninja}~\cite{vanDeurzen:2013saa} and used within the {\Gosam} framework~\cite{refgosam} and \Formcalc~\cite{Nejad:2013ina}, are instead designed to deal with several mass scales and a variety of final states. 
Other versatile codes are \openloops~\cite{Cascioli:2011va} and \recola~\cite{Actis:2012qn}, which build one-loop amplitudes numerically by means of recursion relations applied to Feynman diagrams and off-shell currents, then reduced by means of~{\sc collier}~\cite{Denner:2014gla}. These codes were recently employed for applications involving QCD and EW corrections.
  
The automated computation of physical observables at NLO accuracy, such as cross sections and differential distribution, requires to incorporate the one-loop results for the virtual amplitudes within a Monte Carlo framework (MC). 
In several recent applications, 
the MC provides the possibility of merging multiple NLO parton-level matrix elements with parton showers.
For more details on the MC tools, we refer the reader to the talks of S.~Prestel and F.~Siegert at this Conference.


\section{Higgs boson production in Gluon Fusion with GoSam 2.0}
\label{gosam}

The \gosam~\cite{refgosam} framework combines automated diagram generation and algebraic manipulation~\cite{algebra}
with the integrand reduction techniques and tensorial reduction.
After the generation, the default reduction employed by \gosam\ is the integrand reduction via Laurent expansion provided by {\Ninja}.  Alternatively, the tensorial decomposition provided by {\Golem}~\cite{golem} or the $d$-dimensional integrand reduction as implemented in \SAMURAI\ are also available. 

The code has been employed in numerous applications at NLO QCD accuracy and  
studies of BSM scenarios (see Ref.~\cite{Greiner:2014dfa} for a summary), 
within electroweak calculations~\cite{ewgosam}, 
and recently also within NNLO calculations for the production of real-virtual contributions~\cite{gosamNNLO}.
To achieve these results, \gosam{} has been interfaced within MC tools (a detailed discussion can be found in Ref.~\cite{vanDeurzen:2014uaa}).


As an example of application, we briefly describe the efforts that led to the calculation of NLO QCD corrections to the associated production of a Higgs boson and three jets at the LHC in gluon fusion in the large top-mass limit~\cite{Cullen:2013saa}.

In this limit, the Higgs coupling to gluons mediated by a top-quark loop can be described by an effective operator, leading to new Feynman rules which contains vertices involving the Higgs field and up to four gluons. Such vertices lead to Feynman integral whose rank exceeds the number of denominators. 
A first improvement in \gosam\ needed by this calculation was the upgrade of all reduction algorithms~\cite{Mastrolia:2012bu, vanDeurzen:2013pja, vanDeurzen:2013saa} to cope with additional powers of the integration momentum in the numerator functions. 
As a warm-up process, we tested the algorithm by computing  $pp \to H +2$ jets in gluon fusion~\cite{vanDeurzen:2013rv}.

In order to deal with the complexity level of calculations such as $pp\to H+3$ jets, the {\gosam} code has been further enhanced.  This calculation is indeed challenging both on the side of real-emission contributions and of the virtual corrections, which alone
involve more than ten thousand one-loop Feynman diagrams with up to rank-seven hexagons. The introduction of numerical polarization vectors and the option to sum diagrams sharing the same propagators algebraically during the generation of the code led to an enormous gain in generation time and reduction of code size. Moreover, improvements in the performance have been achieved by exploiting the optimized algebraic manipulation of {\sc Form 4.0}. 
Concerning the reduction, the use of \Ninja\ led to a faster and more stable extraction of all needed coefficients.

An updated analysis appeared in~\cite{Butterworth:2014efa}, which contains new results and distributions for $H+3$ jets at NLO for a set of ATLAS-like cuts and  a comparison with the NLO predictions for $H+2$ jets. Very recently, new phenomenological analyses have been presented~\cite{Greiner:2015jha} which include numerical results for a large variety of observables for both standard cuts and VBF selection cuts.

\section{Beyond NLO}

The Next-to-Next-to-Leading-Order (NNLO) computations are quite far from automation and only a few computations are available for processes at hadron colliders. For a detailed discussion, we refer to the presentation of F.~Petriello in the plenary session.

At one-loop, the advantage of knowing that one complete basis of MIs is formed by scalar 
one-loop functions and the availability of their analytic expression
allowed the community to focus on the development of efficient algorithms  
for the extraction of the coefficients multiplying each MI. 
At higher-loop, a general basis of MIs is not known and they are only identified at the end of the reduction procedure. Moreover, many MIs do not have a known analytic expression and they should be evaluated numerically. 
The multi-loop reduction technique which is most often employed is the well-known Laporta algorithm~\cite{Laporta:2001dd}, based on the solution of algebraic systems of equations obtained through integration-by-parts identities~\cite{Tkachov:1981wb}. 

Recently, new ideas and techniques~\cite{nnlo_new}, along with improved version of known algorithms, are make a huge impact, paving the road to increasingly complex NNLO calculations. The progress in multi-loop calculations and in the computation of Feynman integrals using differential equations are nicely reviewed in the lectures of Refs.~\cite{nnlo_lec}. 
As of now, it is not clear to what extent we will be able to push the available approaches before the computational resources needed become overbearing. In this context, it will be also interesting to observe whether the extensions of integrand-level techniques to higher orders will succeed to provide a reliable alternative option.

\subsection{Integrand-Reduction Techniques Beyond One-Loop}

The reduction at the integrand level is based on the algebraic decomposition of the numerator function ${\cal N}$ of Eq.~(\ref{integr}) in terms of the propagators in the loop, in order to identify before integration the structures that will generate the MIs, as well as terms that will vanish upon integration of the loop momentum but are needed to establish an identity for the integrands. In this approach, the coefficients in front of the MIs can be determined by solving a system of algebraic equations that are obtained by the numerical evaluation of the
numerator of the integrand at explicit values of the loop-variable. The integrand reduction algorithm has been extremely successful for one-loop calculation, and it is the engine within many of the computational tools mentioned in Section~\ref{nlo}.

Extensions beyond one-loop, first proposed in~\cite{Mastrolia:2011pr}, 
have become the topic of several studies~\cite{nnlo_intred}, thus providing a new direction in the study of multi-loop amplitudes. 

Higher-loop techniques require a proper parametrization of the residues at the multi-particle poles~\cite{Mastrolia:2011pr}. 
As in the one-loop case, the parametric form of each polynomial residues is process-independent and can be determined once for all from the corresponding multiple cut. However, at higher loops, the basis of MIs is more complicated and so is the form of the residues.

In Refs.~\cite{Zhang:2012ce, Mastrolia:2012an, Mastrolia:2013kca}, the determination of the residues at the multiple cuts has been systematized as a problem of multivariate polynomial division in algebraic geometry. The use of these techniques proved that the integrand decomposition is applicable not only at one loop, as originally formulated, but at any order in perturbation theory. The shape of the residues is uniquely determined by the on-shell conditions, without any additional constraint. 
Moreover, we presented~\cite{Mastrolia:2012an} a recurrence relation which, independently of the number of loops, leads to the multi-particle pole decomposition of the integrands of the scattering amplitudes. Applications to two-loop Feynman diagrams in QED and QCD showed that the proposed reduction algorithm can be applied to integrands with denominators appearing with arbitrary powers~\cite{Mastrolia:2013kca}.

\section{Summary and Conclusions}

Scattering amplitudes provide an ideal testing ground for many theoretical applications. A better understanding of the mathematical properties of scattering amplitudes indeed allows for the construction of efficient algorithms for their evaluation, and ultimately leads to higher quality theoretical predictions to be used in the experimental analyses at particle colliders.

There is a variety of approaches and numerical tools available for one-loop calculations, which are interfaced with Monte Carlo event generators to provide NLO predictions for processes needed by the LHC experimental collaborations. Just like their tree-level predecessors, these codes allow the user to compute full NLO calculations at the simple effort of 
providing the list of particles and some input parameters. 

Looking ahead, the focus is shifting towards the challenges presented by NNLO calculations, which are considerably more involved. 
While a full automation of NNLO is still not around the corner, there are plenty of activities and studies in the making, and the progress in the field is tangible, both in terms of the development of new ideas and techniques and the completion of new calculations and phenomenological studies. 
%

\paragraph{Acknowledgments} Work supported in part by the National Science Foundation under Grants PHY-1068550 and PHY-1417354. We also acknowledge PSC-CUNY Award No. 67536-00 45.  
%

\begin{thebibliography}{10}
\expandafter\ifx\csname url\endcsname\relax
  \def\url#1{{\tt #1}}\fi
\expandafter\ifx\csname urlprefix\endcsname\relax\def\urlprefix{URL }\fi
\providecommand{\eprint}[2][]{\url{#2}}

\bibitem{review}
Bern Z, Dixon L~J and Kosower D~A 2007 {\em Annals Phys.\/} {\bf 322}
  1587--1634 (\textit{Preprint} \eprint{0704.2798});
Ellis R~K, Kunszt Z, Melnikov K and Zanderighi G 2012 {\em Phys.Rept.\/} {\bf
  518} 141--250 (\textit{Preprint} \eprint{1105.4319});
Ossola G 2014 {\em J.Phys.Conf.Ser.\/} {\bf 523} 012040 (\textit{Preprint}
  \eprint{1310.3214});
Mastrolia P 2015  (\textit{Preprint} \eprint{1507.03226})

\bibitem{Passarino:1978jh}
Passarino G and Veltman M~J~G 1979 {\em Nucl. Phys.\/} {\bf B160} 151;
't~Hooft G and Veltman M 1979 {\em Nucl.Phys.\/} {\bf B153} 365--401

\bibitem{scalars}
van Oldenborgh G 1991 {\em Comput.Phys.Commun.\/} {\bf 66} 1--15;
Hahn T and Perez-Victoria M 1999 {\em Comput.Phys.Commun.\/} {\bf 118} 153--165
  (\textit{Preprint} \eprint{hep-ph/9807565});
Ellis R~K and Zanderighi G 2008 {\em JHEP\/} {\bf 02} 002 (\textit{Preprint}
  \eprint{0712.1851});
van Hameren A 2011 {\em Comput.Phys.Commun.\/} {\bf 182} 2427--2438
  (\textit{Preprint} \eprint{1007.4716});
Cullen G, Guillet J, Heinrich G, Kleinschmidt T, Pilon E {\em et~al.\/} 2011
  {\em Comput.Phys.Commun.\/} {\bf 182} 2276--2284 (\textit{Preprint}
  \eprint{1101.5595})

\bibitem{Bern:1994zx}
Bern Z, Dixon L~J, Dunbar D~C and Kosower D~A 1994 {\em Nucl. Phys.\/} {\bf
  B425} 217--260 (\textit{Preprint} \eprint{hep-ph/9403226})

\bibitem{Britto:2004nc}
Britto R, Cachazo F and Feng B 2005 {\em Nucl. Phys.\/} {\bf B725} 275--305
  (\textit{Preprint} \eprint{hep-th/0412103})

\bibitem{Ossola:2006us}
Ossola G, Papadopoulos C~G and Pittau R 2007 {\em Nucl.Phys.\/} {\bf B763}
  147--169 (\textit{Preprint} \eprint{hep-ph/0609007});
Ossola G, Papadopoulos C~G and Pittau R 2007 {\em JHEP\/} {\bf 0707} 085
  (\textit{Preprint} \eprint{0704.1271});
Ossola G, Papadopoulos C~G and Pittau R 2008 {\em JHEP\/} {\bf 0805} 004
  (\textit{Preprint} \eprint{0802.1876});
  Mastrolia P, Ossola G, Papadopoulos C~G and Pittau R 2008
{\em JHEP} {\bf 06} 030 (\textit{Preprint} \eprint{0803.3964})
  
\bibitem{Giele:2008bc}
Giele W and Zanderighi G 2008 {\em JHEP\/} {\bf 0806} 038 (\textit{Preprint}
  \eprint{0805.2152});   Ellis R, Giele W~T, Kunszt Z and Melnikov K 2009 {\em Nucl.Phys.\/} {\bf B822}
  270--282 (\textit{Preprint} \eprint{0806.3467})

\bibitem{Berger:2008sj}
Berger C, Bern Z, Dixon L, Febres~Cordero F {\em et~al.\/} 2008 {\em
  Phys.Rev.\/} {\bf D78} 036003 (\textit{Preprint} \eprint{0803.4180})

\bibitem{Badger:2012pg}
Badger S, Biedermann B, Uwer P and Yundin V 2013 {\em Comput.Phys.Commun.\/}
  {\bf 184} 1981--1998 (\textit{Preprint} \eprint{1209.0100})

\bibitem{Ossola:2007ax}
Ossola G, Papadopoulos C~G and Pittau R 2008 {\em JHEP\/} {\bf 03} 042
  (\textit{Preprint} \eprint{0711.3596})

\bibitem{Hirschi:2011pa}
Hirschi V, Frederix R, Frixione S, Garzelli M~V {\em et~al.\/} 2011
  {\em JHEP\/} {\bf 1105} 044 (\textit{Preprint} \eprint{1103.0621})

\bibitem{Bevilacqua:2011xh}
Bevilacqua G, Czakon M, Garzelli M, van Hameren A, Kardos A {\em et~al.\/} 2013
  {\em Comput.Phys.Commun.\/} {\bf 184} 986--997 (\textit{Preprint}
  \eprint{1110.1499})

\bibitem{Mastrolia:2010nb}
Mastrolia P, Ossola G, Reiter T and Tramontano F 2010 {\em JHEP\/} {\bf 1008}
  080 (\textit{Preprint} \eprint{1006.0710})

\bibitem{Mastrolia:2012bu}
Mastrolia P, Mirabella E and Peraro T 2012 {\em JHEP\/} {\bf 1206} 095
  (\textit{Preprint} \eprint{1203.0291})

\bibitem{vanDeurzen:2013saa}
van Deurzen H, Luisoni G, Mastrolia P, Mirabella E, Ossola G {\em et~al.\/}
  2014 {\em JHEP\/} {\bf 1403} 115 (\textit{Preprint} \eprint{1312.6678});
Peraro T 2014 {\em Comput.Phys.Commun.\/} {\bf 185} 2771--2797
  (\textit{Preprint} \eprint{1403.1229})

\bibitem{refgosam} 
Cullen G, Greiner N, Heinrich G, Luisoni G, Mastrolia P {\em et~al.\/} 2012
  {\em Eur.Phys.J.\/} {\bf C72} 1889 (\textit{Preprint} \eprint{1111.2034});
Cullen G, van Deurzen H, Greiner N, Heinrich G, Luisoni G {\em et~al.\/} 2014
  {\em Eur.Phys.J.\/} {\bf C74} 3001 (\textit{Preprint} \eprint{1404.7096})

\bibitem{Nejad:2013ina}
Nejad B~C, Hahn T, Lang J~N and Mirabella E 2013  (\textit{Preprint}
  \eprint{1310.0274})

\bibitem{Cascioli:2011va}
Cascioli F, Maierhofer P and Pozzorini S 2012 {\em Phys.Rev.Lett.\/} {\bf 108}
  111601 (\textit{Preprint} \eprint{1111.5206})

\bibitem{Actis:2012qn}
Actis S, Denner A, Hofer L, Scharf A and Uccirati S 2013 {\em JHEP\/} {\bf
  1304} 037 (\textit{Preprint} \eprint{1211.6316})

\bibitem{Denner:2014gla}
Denner A, Dittmaier S and Hofer L 2014 {\em PoS\/} {\bf LL2014} 071
  (\textit{Preprint} \eprint{1407.0087})

\bibitem{algebra}
Nogueira P 1993 {\em J.Comput.Phys.\/} {\bf 105} 279--289;
Vermaseren J~A~M 2000  (\textit{Preprint} \eprint{math-ph/0010025});
Reiter T 2010 {\em Comput.Phys.Commun.\/} {\bf 181} 1301--1331
  (\textit{Preprint} \eprint{0907.3714});
Cullen G, Koch-Janusz M and Reiter T 2011 {\em Comput.Phys.Commun.\/} {\bf 182}
  2368--2387 (\textit{Preprint} \eprint{1008.0803});
Kuipers J, Ueda T, Vermaseren J and Vollinga J 2013 {\em Comput.Phys.Commun.\/}
  {\bf 184} 1453--1467 (\textit{Preprint} \eprint{1203.6543})

\bibitem{golem}
Binoth T, Guillet J~P, Heinrich G, Pilon E and Reiter T 2009 {\em
  Comput.Phys.Commun.\/} {\bf 180} 2317--2330 (\textit{Preprint}
  \eprint{0810.0992});
Heinrich G, Ossola G, Reiter T and Tramontano F 2010 {\em JHEP\/} {\bf 1010}
  105 (\textit{Preprint} \eprint{1008.2441})

\bibitem{Greiner:2014dfa} 
 Greiner N 2014 (\textit{Preprint} \eprint{1410.3237})

\bibitem{ewgosam}
Chiesa M, Montagna G, Barze` L, Moretti M, Nicrosini O {\em et~al.\/} 2013 {\em
  Phys.Rev.Lett.\/} {\bf 111} 121801 (\textit{Preprint} \eprint{1305.6837});
Chiesa M, Greiner N and Tramontano F 2015  (\textit{Preprint}
  \eprint{1507.08579})

%

%


%
%
%

\bibitem{gosamNNLO}
Gao J and Zhu H~X 2014 {\em Phys.Rev.\/} {\bf D90} 114022 (\textit{Preprint}
  \eprint{1408.5150}); 
Gao J and Zhu H~X 2014 
{\em Phys.Rev.Lett.\/} {\bf 113} 262001
  (\textit{Preprint} \eprint{1410.3165});
Del~Duca V, Duhr C, Somogyi G, Tramontano F and Trocsanyi Z 2015
  (\textit{Preprint} \eprint{1501.07226})

\bibitem{vanDeurzen:2014uaa}
van Deurzen H, Greiner N, Heinrich G, Luisoni G {\em et~al.\/}
  2014 {\em PoS\/} {\bf LL2014} 021 (\textit{Preprint} \eprint{1407.0922})


%
%
%
%

\bibitem{Cullen:2013saa}
Cullen G, van Deurzen H, Greiner N, Luisoni G, Mastrolia P {\em et~al.\/} 2013
  {\em Phys.Rev.Lett.\/} {\bf 111} 131801 (\textit{Preprint}
  \eprint{1307.4737})

\bibitem{vanDeurzen:2013pja}
van Deurzen H 2013 {\em Acta Phys.Polon.\/} {\bf B44} 2223--2230;
Guillet J~P, Heinrich G and von Soden-Fraunhofen J~F 2014 {\em Comput. Phys.
  Commun.\/} {\bf 185} 1828--1834 (\textit{Preprint} \eprint{1312.3887})

\bibitem{vanDeurzen:2013rv}
van Deurzen H, Greiner N, Luisoni G, Mastrolia P, Mirabella E {\em et~al.\/}
  2013 {\em Phys.Lett.\/} {\bf B721} 74--81 (\textit{Preprint}
  \eprint{1301.0493})


\bibitem{Butterworth:2014efa}
Butterworth J, Dissertori G, Dittmaier S, de~Florian D, Glover N {\em et~al.\/}
  2014  (\textit{Preprint} \eprint{1405.1067})

\bibitem{Greiner:2015jha}
Greiner N, Hoeche S, Luisoni G, Schonherr M, Winter J~C and Yundin V 2015
  (\textit{Preprint} \eprint{1506.01016})

\bibitem{Laporta:2001dd}
Laporta S 2000 {\em Int.J.Mod.Phys.\/} {\bf A15} 5087--5159 (\textit{Preprint}
  \eprint{hep-ph/0102033})

\bibitem{Tkachov:1981wb}
Tkachov F 1981 {\em Phys.Lett.\/} {\bf B100} 65--68

\bibitem{nnlo_new}
Henn J~M 2013 {\em Phys.Rev.Lett.\/} {\bf 110} 251601 (\textit{Preprint}
  \eprint{1304.1806});
Argeri M, Di~Vita S, Mastrolia P, Mirabella E, Schlenk J {\em et~al.\/} 2014
  {\em JHEP\/} {\bf 1403} 082 (\textit{Preprint} \eprint{1401.2979});
Papadopoulos C~G 2014 {\em JHEP\/} {\bf 1407} 088 (\textit{Preprint}
  \eprint{1401.6057});
Gehrmann T, von Manteuffel A, Tancredi L and Weihs E 2014 {\em JHEP\/} {\bf
  1406} 032 (\textit{Preprint} \eprint{1404.4853});
Lee R~N 2014  (\textit{Preprint} \eprint{1411.0911})

\bibitem{nnlo_lec}
Duhr C 2014  (\textit{Preprint} \eprint{1411.7538});
Henn J~M 2015 {\em J. Phys.\/} {\bf A48} 153001 (\textit{Preprint}
  \eprint{1412.2296})

\bibitem{Mastrolia:2011pr}
Mastrolia P and Ossola G 2011 {\em JHEP\/} {\bf 1111} 014 (\textit{Preprint}
  \eprint{1107.6041})

\bibitem{nnlo_intred}
Badger S, Frellesvig H and Zhang Y 2012 {\em JHEP\/} {\bf 1204} 055
  (\textit{Preprint} \eprint{1202.2019});
Kleiss R~H, Malamos I, Papadopoulos C~G and Verheyen R 2012 {\em JHEP\/} {\bf
  1212} 038 (\textit{Preprint} \eprint{1206.4180});
Badger S, Frellesvig H and Zhang Y 2012 {\em JHEP\/} {\bf 1208} 065
  (\textit{Preprint} \eprint{1207.2976});
Feng B and Huang R 2013 {\em JHEP\/} {\bf 1302} 117 (\textit{Preprint}
  \eprint{1209.3747});
Mastrolia P, Mirabella E, Ossola G and Peraro T 2013 {\em Phys.Rev.\/} {\bf
  D87} 085026 (\textit{Preprint} \eprint{1209.4319})

\bibitem{Zhang:2012ce}
Zhang Y 2012 {\em JHEP\/} {\bf 1209} 042 (\textit{Preprint} \eprint{1205.5707})

\bibitem{Mastrolia:2012an}
Mastrolia P, Mirabella E, Ossola G and Peraro T 2012 {\em Phys.Lett.\/} {\bf
  B718} 173--177 (\textit{Preprint} \eprint{1205.7087})

\bibitem{Mastrolia:2013kca}
Mastrolia P, Mirabella E, Ossola G and Peraro T 2013 {\em Phys.Lett.\/} {\bf
  B727} 532--535 (\textit{Preprint} \eprint{1307.5832})

\end{thebibliography}

{\small

\providecommand{\newblock}{}

}



\end{document}